# AUTOMATIC PLANETARY DEFENSE : DEFLECTING NEOs BY MISSILES SHOT FROM L1 AND L3 (EARTH-MOON)

## Claudio Maccone

Member of the International Academy of Astronautics
Address: Via Martorelli, 43 - Torino (TO) 10155 - Italy
Phone: +39-011-7180-313. Fax: +39-011-7180-012.
URL: http://www.rli.it/  - E-mail: clmaccon@libero.it

**ABSTRACT.** We develop the mathematical theory for an automatic, space-based system to deflect NEOs by virtue of missiles shot from the Earth-Moon L1 and L3 Lagrangian Points.

A patent application has been filed for the relevant code dubbed AsterOFF (=Asteroids OFF !). This code was already implemented, and a copyright for it was registered.

In a paper published in Acta Astronautica, Vol. 50, No. 3, pp. 185-199 (2002), this author proved mathematically the following theorem (hereafter called the "confocal conics theorem"):

"Within the sphere of influence of the Earth, any NEO could be hit by a missile at just an angle of 90 degrees, was the missile shot from the Lagrangian Points L1 or L3 of the Earth-Moon system, rather than from the surface of the Earth". As a consequence, the hitting missile would have to move along a "confocal ellipse" (centered at the Earth) uniquely determined by the NEO's incoming hyperbola.

Based on the above theorem, the author further shows in this paper that:

1) The proposed defense system would be ideal to deflect NEOs that are small, i.e. less than one kilometer in diameter. Small NEOs are just the most difficult ones to be detected early enough and to such an orbital accuracy to be positively sure that they are indeed hazardous.
2) The traditional theory of Keplerian orbits can successfully be applied to get an excellent first-order approximation of the (otherwise unknown) mathematical formulae of the energy-momentum requested to achieve the NEO deflection. Many engineering details about the missiles shot from L1 and L3, however, still have to be implemented into our simulations, partly because they are classified.
3) Was one missile not enough to deflect the NEO completely, it is a great advantage of the "confocal conics" used here that the new, slightly deflected NEO's hyperbola would certainly be hit at nearly 90 degrees by another and slightly more eccentric elliptical missile trajectory. A sufficient number of missiles could thus be launched in a sequence from the Earth-Moon Lagrangian points L1 and L3 with the result that the SUM of all these small and repeated deflections will finally throw the NEO off its collision hyperbola with the Earth.

## 1. INTRODUCTION

An innovative paper (ref. [1]) was published in February 2002 by this author. It laid the mathematical foundations of a new concept of Planetary Defense of the Earth against hazardous asteroids and comets. The guidelines of this new vision are:

1) It is hard to deflect something that's coming right at you if you keep staying on the surface of the Earth. Thus, it is unreasonable to do any effective Planetary Defense from the surface of the Earth, and we must do it from space instead.
2) Where in space? The nearest two Lagrangian Points L1 and L3 of the Earth-Moon system are the two most obvious locations since they keep the same distance from Earth at all times. We thus need to create missile bases there ready to be shot against hazardous asteroids (as hazardous asteroids and comets will briefly be called hereafter). The two Earth-Moon Lagrangian Points L1 and L3 not only are nearly-stable points (i.e. any missile base there won't "fly away" from Earth into space). More important still, the trajectories of all missiles shot from there will intercept any incoming asteroid at an angle of 90 degrees if shot along ellipses confocal to the asteroid's hyperbola (confocal conics theorem). This "orthogonal interception property" automatically insures the maximum sideway momentum transfer from the missile to the asteroid. In addition, a large steel basket open up the missile head would help pushing the asteroid sideways, especially if the asteroid is small.

3) Not just that. A further advantage of the confocal conics theorem (used in ref. [1] to prove mathematically that the missile-asteroid collision would always occurs at an angle of 90 degrees) has one more consequence of exceptional practical importance that we call the "repeated deflection capability". Here is what this is: suppose that one missile-asteroid collision occurs, but the deflection in the asteroid's path is not large enough to bring it off its collision course with the Earth. Since the Earth always lies at the common focus of both the asteroid and missiles trajectories, the confocal conics theorem insures that all the elliptical trajectories of subsequent missiles are orthogonal to whatever hyperbolic trajectory the asteroid may have. This basic result means that we may shoot more the one missile in a sequence and have the asteroid deflected from one hyperbola to the next more eccentric one, and so on and so on for as many times as it may be needed until we finally push the asteroid off its collision course with the Earth. We like to call this result the "cumulative effect" of the repeated interception capability, and regard this "march of the dimes" of many smaller deflections totaling up into one, larger deflection as the key to saving Humankind from the impact.

This paper is devoted to a much more comprehensive mathematical description of the principles stated in ref. [1]. But this introduction to the subject would not be complete without mentioning the several "popular descriptions" of the author's paper [1] that were published all over the world after February 2002. Most of these popular descriptions may be downloaded from the Cambridge Conference Correspondence web site http://abob.libs.uga.edu/bobk/ccc/cc021402.html, but more similar, popular descriptions are possibly unknown to this author. The known-to-him popular summaries and comments were given by, respectively:

1) The popular science magazine "New Scientist" in an article titled "INCOMING! TO DEFLECT AN ASTEROID, CHOOSE YOUR SHOT CAREFULLY" by Eugenie Samuel. It is now available at the Cambridge Conference Correspondence site mentioned above. The author of this paper is grateful to Eugenie Samuel for her timely interest in his work and for her phone interview.
2) By "Ananova" in an article titled "Astronomer says space rockets may be Earth's best defense", available at the web site http://www.ananova.com/news/story/sm_520279.html .
3) By the "Sunday Herald", a short summary within a review article now available at the site http://www.sundayherald.com/26551 .
4) The senior science writer Robert Roy Britt wrote a popular description in the language of boxers! It is titled "Space-Based Missile Defense Needed to Thwart Asteroid Attacks" and is at http://www.space.com/scienceastronomy/solarsystem/deflection_asteroids_020214.html . . This description also is at the NASA-JPL archive http://neo.jpl.nasa.gov/news_archives0206.html look for the date of February 14, 2002. The author of this paper is grateful to Robert Roy Britt for his neat article.
5) One more synthetic description is found at the British National Space Center web site http://www.nssc.co.uk/spacenow/spacenewsitem.asp?N=402&R=1 .
6) In German, a short description was later given in an article titled "Wie man Asteroiden abwehrt", available at the site http://www.pm-magazin.de/de/wissensnews/wn_id171.htm .
7) The author of this paper apologizes for not mentioning possible other "popular" web sites currently unknown to him.

Finally, it must be pointed out that this author's plan for Planetary Defense also has deep political and military implications, as it always happened for all new breakthroughs in the history of astronautics. To summarize them, we would like to use the same words that the British impact-expert Benny Peiser is reported (in Robert Roy Britt's article described above at point 4) ) to have said after having read the author's first paper [1]: "Peiser figures that a plan like Maccone's would have to be led by the U.S. Military, which already sees space as a necessary strategic outpost". The current response to terrorism, Peiser said, "has led the U.S. to significantly increase the budget for space-based defense paraphernalia which inadvertently enhances the prospects for advanced planetary defense technologies."

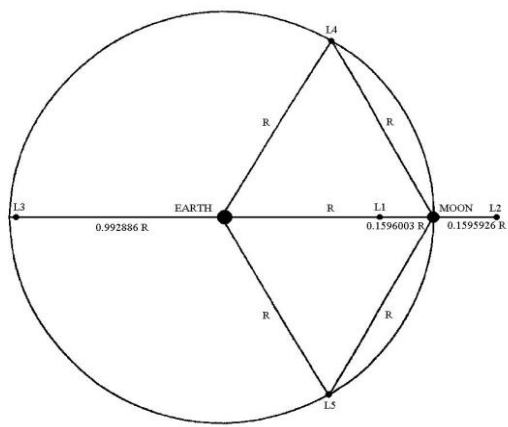

**Figure 1. The five Lagrangian points of the Earth-Moon system and their distances from Earth and Moon expressed in terms of R, the Earth-Moon distance (supposing the Moon orbit circular, in the first approximation).**

## 2. THE 5 LAGRANGIAN POINTS OF THE EARTH-MOON SYSTEM: A SUMMARY

To pave the way to the new mathematical results proven in this paper for the first time (see Sections 4 and higher), it may not be a waste of time to summarize the basic equations already proven by this author in his first two papers on the subject, refs. [1] and [2]. In ref. [2] an extension of the results proven in ref. [1] was given, in that the equation of the missile ellipses to be shot from the Earth-Moon L1 point were given there for the first time.

Let us start by reviewing where the five Lagrangian points of the Earth-Moon system are located, even if this is well-known material to astrophysicists and space scientists alike (see, for instance, the excellent paper [3] written in 1987 by the American astrodynamicist David W. Dunham).

Historians of science tell that the year was 1772 when mathematician Joseph Louis Lagrange (1736-1813) won the prize of the Paris Academy of Sciences for having demonstrated that five positions of net zero-force exist in a rotating two-body gravity field (refs. [4] and [5] give the full account). Out of these five "Lagrangian Points" (also called "Libration Points"), three are situated on the line joining the two massive bodies, and are nowadays called "colinear points", or L1, L2 and L3, as shown in Figure 1 for the Earth-Moon system. The other two (called L4 and L5) form equilateral triangles with the two massive bodies, and so are called "triangular points".

The locations of the three colinear points L1, L2 and L3 are found as the real roots of an algebraic equation of the fifth degree, originally due to Lagrange, that can only be solved by resorting to Taylor series expansions. Fortunately, the relevant three different Taylor expansions converge rapidly, so one may take into account just three terms in each Taylor expansion to get approximations that are quite satisfactory numerically. Here we just confine ourselves to stating that, assuming for the Earth-Moon distance the numerical value of R = 384,401 km, then:

1) The distance between the Moon and the Lagrangian point L1 equals 0.1596003*R, that is 61350.317208 km. Consequently the Earth-to-L1 distance equals 0.8403997*R, that is 323050.482792 km.
2) The distance between the Moon and the Lagrangian point L2 equals 0.1595926*R, that is 61347,568938 km.
3) The distance between the Earth and the Lagrangian point L3 equals 0.992886*R, that is 381666.370650 km.

## 3. APPLYING THE CONFOCAL CONICS THEOREM TO PLANETARY DEFENSE: MATHEMATICAL RESULTS ALREADY PUBLISHED IN REFS. [1] AND [2]

The application of the confocal conics theorem to the deflection of asteroids already was published refs. [1], [2] and [3], along with many of the relevant mathematical proofs. This section is then devoted to a non-mathematical description of ***confocal conics*** as the ***best trajectories*** for deflecting dangerous asteroids by virtue of missiles launched from either of the two colinear Lagrangian points L1 and L3. The triangular points L4 and L5 are excluded from this theory because the trajectory of missiles launched from them is not planar, and so much more complicated. Also, the Lagrangian points L1 and L2 of the Sun-Earth system are excluded from the following considerations, inasmuch as they would require the theory of perturbations. In conclusion, we just consider the ***Planetary Defense from the nearest two Lagrangian points, L1 and L3.***

Consider defined two families of confocal conics, i.e. of conics that have the same foci, as shown in Figure 2.

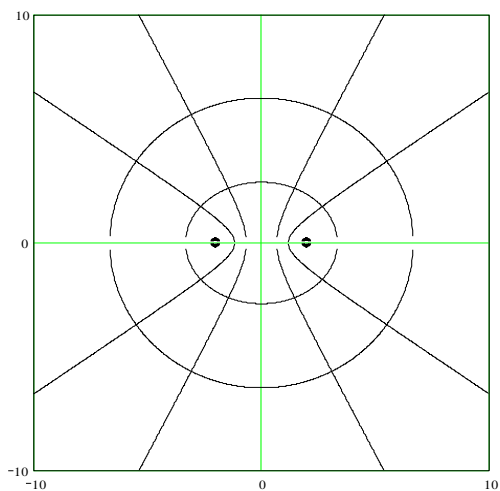

**Figure 2. The two families of $\infty^1$ confocal ellipses and $\infty^1$ confocal hyperbolas.**

One is the family of $\infty^1$ confocal ellipses, and the other is the family of $\infty^1$ confocal hyperbolas. In fact, each one of the confocal ellipses has a different value of both $a_{ell}$ and $e_{ell}$, but the latters' product always equals the constant value $c$ of the distance of the foci from the origin of the axes. Then , the "great" property of confocal conics is that the any pair of different confocal conics, namely one ellipse and one hyperbola, always intersect each other at angles of 90°. In other words, the family of ellipses and the family of hyperbolas, form two families of **orthogonal trajectories.** This is a well-known result proven in any textbook of elementary analytical geometry.

Is all this good for deflecting dangerous asteroids ?

Yes - is our answer - and here follows the sequence of logical steps leading towards the application of confocal conics as the **_best_** trajectories for missiles shot from the Lagrangian points :

1) Consider only one of the two focuses (or "foci", in Latin), namely the one "on the left". Imagine the Earth is there. Then both the confocal ellipse and the confocal hyperbola are physical trajectories (paths) that moving bodies around the Earth follow naturally, because this is just Kepler's First Law. But, which body is following which path ?
2) A dangerous asteroid or comet arrives from infinity, namely from outside the sphere of influence of the Earth. So, it can only follow a *hyperbolic* trajectory with respect to the Earth, with the focus of this hyperbola located just at the center of the Earth (at least in the first approximation, to which we confine ourselves here). Also, the incoming asteroid or comet is to be regarded as "dangerous" only if its path crosses the Earth, namely if the perigee of its hyperbolic trajectory is smaller than the Earth radius: $c_{hyp} \leq R_{Earth}$.
3) What is the counterpart of the ellipse confocal to the asteroid's hyperbola ? Our answer is: the confocal ellipse is the physical trajectory of a missile launched against the incoming dangerous asteroid from any point in space, but better from the two colinear Lagrangian points, L1 and L3, located each on one side of the Earth "for better defense". Points L4 and L5 could also be used, but the relevant missile's orbit calculations would be more involved as three-dimensional. Notice also that L2 is to be excluded from becoming a missile base because not visible from the Earth (the Moon hides it). The selection of the colinear Lagrangian points L1 and L3 as space bases for missiles is now self-evident: they ensure the **cylindrical symmetry of the problem around the Earth-Moon axis**. So, the direction in space from which the asteroid is arriving towards the Earth becomes *irrelevant* (at least in this first-order approximation): we will just be studying confocal orbits in the plane passing through the Earth-Moon axis and the asteroid. Additionally, the merit of all the Lagrangian points is that they are "fixed" in the Earth-Moon system, in that they keep their positions unaltered with respect to the Earth and the Moon at all times.
4) But being confocal, the missile's ellipse is automatically **orthogonal** to the asteroid's hyperbola, meaning that the collision of the missile with the asteroids always occurs at a right angle with the asteroid's path. This is really **_the best_** we can hope for in order to deflect the asteroid, since the missile's **_full_** momentum is then transferred to the asteroid *sidewise*.
5) Finally, if one missile fails to deflect the asteroid's path in a sufficient amount, we can always send one or more missiles again along the new ellipse that is confocal to the new and slightly deflected asteroid's hyperbolic path. This is because confocal conics are actually two *families* of $\infty^1$ trajectories. So, once again, the mathematical representation of the trajectories in the game by virtue of **confocal** conics matches perfectly with the physical problem of diverting dangerous asteroids and comets!

Having ascertained the above five facts, we now face a mathematical problem: given the trajectory of the incoming asteroid, that is given its confocal hyperbola, can the relevant confocal ellipse departing from L3 or L1 be determined uniquely ?

Yes - is the answer - as it was proven in ref. [1], and we shall not re-prove this basic fact here. We just confine ourselves to state that the eccentricity of the ellipse of the missile shot from L3 is

$$e_{ell\_from\_L3} = \frac{\sqrt{R_{L3}^2 + 4R_{L3}\,a_{hyp}\,e_{hyp}\cos(\omega_{hyp}) + 4a_{hyp}^2\,e_{hyp}^2} - R_{L3}}{2[R_{L3}\cos(\omega_{hyp}) + a_{hyp}\,e_{hyp}]}.$$
(3.1)

The subscript "ell_from_L3" in the last equation replaced the simple "ell" to make it clear that this is the ellipse of all missiles shot from the L3 Lagrangian Point, rather than from the L1 one. The semi-major axis of the same ellipse is

$$a_{ell\_from\_L3} = \frac{2[R_{L3}\cos(\omega_{hyp}) + a_{hyp}\,e_{hyp}]\,a_{hyp}\,e_{hyp}}{\sqrt{R_{L3}^2 + 4R_{L3}\,a_{hyp}\,e_{hyp}\cos(\omega_{hyp}) + 4a_{hyp}^2\,e_{hyp}^2} - R_{L3}}.$$
(3.2)

and the problem is solved.

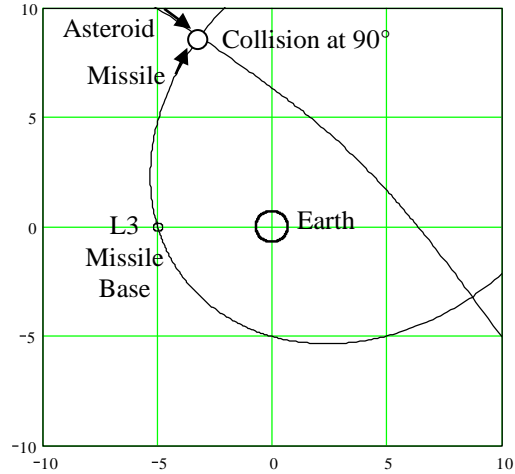

**Figure 3. Asteroid missing the Earth, and elliptical path of the missile shot against it from L3. The orthogonality of the two paths at their collision point is evident (the scale of this drawing is arbitrary).**

Figure 3 shows a first numerical example (in arbitrary units not to scale with the actual Earth-Moon system values): the incoming asteroid is missing the Earth (represented by the larger circle located at the origin), but is of course deflected along a hyperbola. Then the missile shot from the Lagrangian point L3 (the smaller circle on the left) can hit the asteroid before it approaches the Earth, colliding with it at an orthogonal angle.

We would like to complete this section with a small summary followed by a key remark.

In this section we have proven that the missile ellipse is completely determined by the asteroid's hyperbola. This means that:

1) Determining the asteroid's hyperbola is the first goal that must be achieved. Currently (August 2004) many potentially hazardous asteroids' hyperbolae (with respect to the Earth) are given at the NASA-JPL NEO web site http://neo.jpl.nasa.gov/orbits/.
2) Once the asteroid's hyperbola is known, the missile ellipse is known too because the latter is a mathematical consequence of the former.
3) But then... an ***automatic defense system*** (hence possibly unmanned) could be built at L1 and L3, ready to shoot ***automatically*** whenever an asteroid would come along. This is our ***key remark***.

## 4. POINT IN SPACE WHERE THE MISSILE INTERCEPTS THE ASTEROID

The point of intercept between missile and asteroid is where they collide. In mathematical terms, this is where the two radii in the equations of the ellipse and of the hyperbola equal each other. One might be tempted to "invert" the last equation by invoking the acos(...) function, but that would make us "loose" either of the two resulting roots on the way. The correct procedure, having set

$$A = \frac{a_{hyp}(e_{hyp}^2 - 1) - a_{ell}(1 - e_{ell}^2)}{e_{ell}\,a_{hyp}(e_{hyp}^2 - 1) + a_{ell}(1 - e_{ell}^2)e_{hyp}}, \quad (4.1)$$

yields

$$\phi_1 = 2\operatorname{atan}\left(\frac{\sqrt{1-A^2} + \sin(\omega_{hyp})}{A + \cos(\omega_{hyp})}\right) + m \cdot 2\pi \quad (4.2)$$

$$\phi_2 = -2\operatorname{atan}\left(\frac{\sqrt{1-A^2} - \sin(\omega_{hyp})}{A + \cos(\omega_{hyp})}\right) + n \cdot 2\pi. \quad (4.3)$$

Eqs. (4.2) and (4.3) (where *m, n* =1,2,3...) yield the different true anomalies $\phi_1$ and $\phi_2$ of the two points where the asteroid's hyperbola is intercepted by the missile's confocal ellipse. Of course, only ***one*** of these two points will actually used to deflect the hazardous asteroid: the point closer to the asteroid's incoming direction.

As for the distance between the origin (Earth's center) and either of the two intercepts, after some reductions, is found to be given by

$$r_{intercept} = \frac{e_{ell} a_{hyp}\left(e_{hyp}^2 - 1\right) + a_{ell} e_{hyp}\left(1 - e_{ell}^2\right)}{e_{ell} + e_{hyp}}. \quad (4.4)$$

Once again we would like to point out that, upon replacing the missile ellipse's a and e (eqs. (3.1) and (3.2)) into eqs. (4.2) and (4.3), the polar coordinates of the point of intercept become known functions of the asteroid's hyperbolic trajectory only. This actually means that *the Planetary Defense from L1 and L3 could be made fully automatic.* In other words: as soon as an asteroid (especially a *small* asteroid) arrives, an *automatic* missile launch system shoots the missiles from L1 or L3, with possibly no human having to spend long times at L1 or L3 or having to rush there, wasting precious time!

## 5. SHOT ANGLE AND SHOT SPEED FOR MISSILES LAUNCHED FROM L1 AND L3

The space bases for missiles at L1 and L3 will have to be carefully designed in order to optimize for the missile launches. Thus, knowledge of the missiles shot angle speed will be a central feature of this design work. We give equations for these two basic features in this section.

The "missile shot angle", $\vartheta_{missile\_shot}$, we define as the angle in between the Earth-Moon axis and the tangent to the missile's ellipse at the launch base (either at L3 or at L1). We assume this angle to be positive when computed counterclockwise. Easy calculations that we do not describe here yield

$$\vartheta_{missile\_shot\_from\_L3} = \mathrm{atan}\left(\frac{\left.\frac{dy_{ell}}{d\phi}\right|_{\phi=\pi}}{\left.\frac{dx_{ell}}{d\phi}\right|_{\phi=\pi}}\right) \quad (5.1)$$

$$\vartheta_{missile\_shot\_from\_L1} = \mathrm{atan}\left(\frac{\left.\frac{dy_{ell}}{d\phi}\right|_{\phi=0}}{\left.\frac{dx_{ell}}{d\phi}\right|_{\phi=0}}\right). \quad (5.2)$$

One more key feature for the L3 and L1 space base design is the missile (initial) shot speed, defined as the missile speed along the ellipse computed at either L3 or L1. We approximate by forgetting about the transient phase of the launch, during which the missile speed increases from zero up to the (initial) shot speed $v_0$. Then, the energy integral along the ellipse shot from L3 yields at once

$$v_{0\_from\_L3} = \sqrt{GM_{Earth}}\sqrt{\frac{2}{R_{L3}} - \frac{1}{a_{ell\_from\_L3}}} \quad (5.3)$$

where all variables appearing in the right hand sides are known because of (3.2). Similarly, as for the missile ellipse shot from L1 one gets

$$v_{0\_from\_L1} = \sqrt{GM_{Earth}}\sqrt{\frac{2}{R_{L1}} - \frac{1}{a_{ell\_from\_L1}}} \quad (5.4)$$

where the analogous to (3.2) must be replaced. In conclusion, the point to notice is that the missile launch speed depends on the asteroid's incoming direction and speed, of course.

## 6. TIME TO DOOMSDAY, AND TIME FOR MISSILES TO AVOID IT!

The Planetary Defense Theory described in this paper is Keplerian. One may thus immediately apply well-known Keplerian results. For instance, of paramount importance in Planetary Defense is the "Time to Doomsday", i.e. the time left between the first intercept of the asteroid by a missile shot from L3 and the instant the asteroid would crash against the Earth. This is the time still left had there been no intercept at all, or had the interception been insufficient to deflect to asteroid from its collision course with the Earth. Let F be, the hyperbolic orbits analogous of the eccentric anomaly E. Then F is expressed in terms of $\phi$ by

$$F(\phi) = 2\,\mathrm{atanh}\left(\sqrt{\frac{e_{hyp}-1}{e_{hyp}+1}}\tan\left(\frac{\phi}{2}\right)\right). \quad (6.1)$$

The time spent along any hyperbolic orbit, and in between the true anomalies $\phi_1$ and $\phi_2$, is then given by the hyperbolic analogue of Kepler's equation

$$t_2 - t_1 = \frac{(a_{hyp})^{\frac{3}{2}}}{\sqrt{GM_{Earth}}} \cdot$$
$$\cdot \left[\left(e_{hyp}\sinh(F(\phi_2)) - F(\phi_2)\right) - \left(e_{hyp}\sinh(F(\phi_1)) - F(\phi_1)\right)\right].$$
$$(6.2)$$

This is just the time between any instant $t_1$ to Doomsday if one identifies the final instant $t_2$ with the time of the asteroid's passage at the perigee, described mathematically by the condition

$$\phi_2 = \omega_{hyp} \quad (6.3).$$

Knowing how much time we are left "to do something" would help us "doing something" by shooting missiles in advance from L1 or L3. How long in advance? To answer this question it is vital for us to know how long it would take for missiles shot from L1 or L3 to reach the asteroid along the missiles elliptical trajectory. And this time is found by virtue of the Kepler equation for the missile ellipse

$$t_2 - t_1 = \frac{(a_{ell})^{\frac{3}{2}}}{\sqrt{G M_{Earth}}} \cdot$$
$$\cdot [(E(\phi_2) - e_{ell} \sin(E(\phi_2))) - (E(\phi_1) - e_{ell} \sin(E(\phi_1)))]$$
(6.4)

where:

1) The ellipse semi-major axis $a_{ell}$ is given by (3.2) for L3 or the equivalent for L1;
2) The eccentric anomalies of the L1 and L3 points just equal the relevant true anomalies (0 and π, respectively) since, just for these two special numerical values, true anomalies and eccentric anomalies are the same.

In general, the missile time along its own ellipse is higher than the asteroid's time along its own hyperbola. Thus, missile shots must be planned in advance as much as possible. In addition, for very large asteroids, like big comets or asteroids several km in side, it would be more appropriate to take into account the reduced mass of the Earth+Asteroid system in all of the above equations, rather than just the mass of the Earth alone. Another "little flaw" that should be taken into account is the possibility that the asteroid crossed the sphere of influence of the Moon on its way to get very close to the Earth. This would change the shape of the otherwise purely hyperbolic orbit of the asteroid into a more complicated, that could be found in an approximate fashion by using the patched conics method. A final, very general remark on the times of flight just described. One may say that it would take up to some 20 days or so for both the missiles to reach their target (and deflect it) and a shorter time for the asteroid to reach the Earth. ***Our Confocal Planetary Defense System is thus a "Last Ditch Defense".***

## 7. ASTEROID'S ORTHOGONAL DEFLECTION: DETAILED MATHEMATICAL THEORY

This is the most important section in this paper. The mathematical theory of asteroid deflection at 90º by letting the asteroid be hit by a missile is presented in this section in detail. A code written by this author in the "MathCad" environment was basic to check numerically many mathematical details that could hardly be understood in pure analytical terms only.

We are going to solve the following problem:

given the three orbital elements of the incoming asteroid hyperbola, $(a_1, \ e_1, \ \omega_1)$,

find the corresponding three orbital elements $(a_2, \ e_2, \ \omega_2)$ of the new, deflected asteroid hyperbola.

To solve this problem, please consider the diagram shown in Figure 4. An asteroid arrives "from the left", or, more appropriately, from the side opposite to the position of the Moon. It is hit at 90º by a missile shot from the missile space base located at L3. Let us then call:

1) "Collision" the crash of the missile against the asteroid at intercept. This crash is supposed to last a very small time when compared to the motions of both the asteroid and missiles: thus, the collision is supposed to be "instantaneous". Under these circumstances, we can apply the law of conservation of the linear momentum at the point of intercept. This we will do in this Section. We already know from the results given in section 4 that the polar coordinates $(r_{\text{intercept}}, \ \phi_{\text{intercept}})$ of the intercept point are completely determined by the incoming asteroid hyperbola elements $(a_1, \ e_1, \ \omega_1)$ by virtue of eqs. (4.2), (4.3) and (4.4). We just add that, in order to simplify the notations, we are going to put the suffix "1" to all the mathematical variables *before* the collision, and put the suffix "2" to all the mathematical variables *after* the collision. In other words, all mechanical variables relevant to the incoming asteroid's hyperbola are going to be denoted by the subscript 1, and all variables related to the outgoing (or deflected) asteroid's hyperbola are going to be denoted by the subscript 2.

2) Denote by $\vec{V_1}$ the asteroid velocity just before the collision. This $\vec{V_1}$ is a vector tangent to the asteroid's hyperbola, and is larger in module than the speed of anything moving along any ellipse. Typically, observational data about the most recent "near misses" by NEOs (= Near Earth Objects), listed at the NASA-JPL NEO web site http://neo.jpl.nasa.gov/neo/close.html, give values of $V_1$ ranging roughly in between 1.5 km/sec and 40 km/sec. These numbers are much higher than the missile speeds along ellipses departing from L3 and ranging in between 1.022 km/sec and 1.445 km/sec only, as pointed out at point 3) hereafter. The mathematical expression of $V_1$ in terms if the incoming asteroid hyperbola elements $(a_1, e_1, \omega_1)$ and of the intercept point $(r_{intercept}, \phi_{intercept})$ follows at once from the energy integral applied to the incoming asteroid hyperbola and solved for $V_1$

$$V_1 = \sqrt{GM_{Earth}} \sqrt{\frac{2}{r_{intercept}} + \frac{1}{a_1}} \qquad (7.1)$$

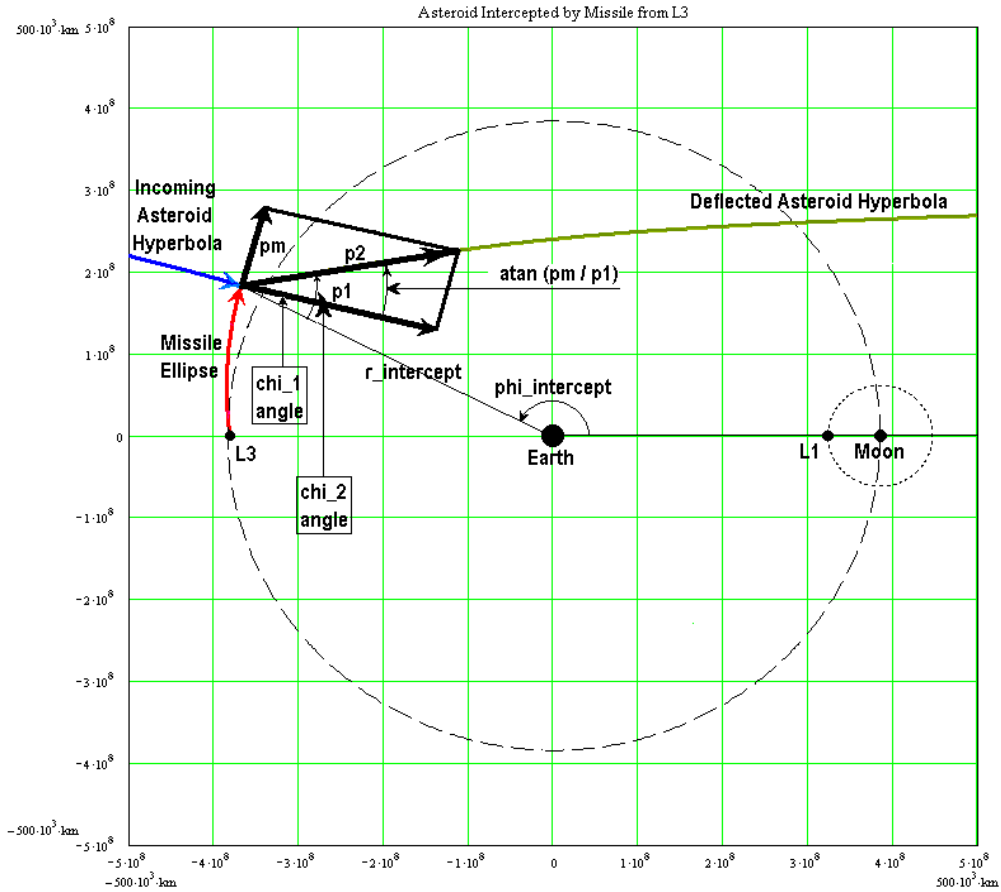

**Figure 4. How the Orthogonal Deflection works for an Asteroid arriving towards the Earth "from the side opposite to the Moon". Before it reaches the Earth, a missile shot from L3 hits the asteroid at 90 degrees. The linear momentum is conserved during the collision between the missile and the asteroid. But the asteroids' orbital angular momentum is NOT conserved, NOR is conserved the asteroid's orbital total energy. Thus, the initial asteroid's hyperbola is changed into a new, different DEFLECTED HYPERBOLA, the three parameters of which (a, e, ω) are derived analytically in this Section.**

3) $V_{missile}$ the missile speed just before the collision. This speed may be regarded as the sum of the two different contributions, as now described.
4) $V_{ell}$, the missile speed along any ellipse from L3 given by eq. (5.3) with $v_{0\_from\_L3}$ replaced by $V_{ell}$, and $R_{L3}$ replaced by $r_{intrceptl}$ of (4.4). This shows that the numerical values for $V_{ell}$ only may range in between 1.022 km/sec (speed along a circular orbit around the Earth through L3) and 1.445 km/sec (the parabolic speed for a missile shot from L3).
5) In order to increase the amount of deflection of the incoming asteroid, however, one might well impart a boost to the missile just seconds before it crashed into the asteroid. An infrared sensor on top of the missile might direct it straight into the asteroid. Since the ellipse is almost a straight line just before the collision, this extra-boost speed adds to the natural speed along the ellipse, $V_{ell}$, with the result that the overall missile speed at the collision instant with asteroid is given by

$$\boxed{V_{missile} = V_{ell} + V_{boost}} \quad (7.2)$$

***This innocent-looking formula will later prove to be of paramount importance for the selection of the deflecting missiles and, thus, for the final characterization of any future Confocal Planetary Defense System.*** Very roughly, $V_{boost}$ might reach up to something like 9 km/sec, so that the overall missile speed at collision, $V_{missile}$, could total something like 10 km/sec.

Many more important equations can be derived by the simple inspection of Figure 4.

6) The first, basic inference stems out of the Pythagorean theorem applied to the rectangular triangle spanned by the orthogonal vectors $V_{missile}$ and $V_1$. Multiplication of each such velocity times the relevant moving mass yields the corresponding linear momentum denoted by *p* with the relevant suffix. We thus have by definition

$$\begin{cases} p_{missile} = M_{missile} V_{missile} \\ p_1 = M_{asteroid} V_1 \\ p_2 = (M_{asteroid} + M_{missile}) V_2 \end{cases} \quad (7.3)$$

Then, the linear momentum conservation law yields at once

$$\vec{p}_2 = \vec{p}_1 + \vec{p}_{missile}. \quad (7.4)$$

This vector equation, applied to the rectangular triangle, yields

$$p_2^2 = p_1^2 + p_{missile}^2 \quad (7.5)$$

that is

$$(M_{asteroid} + M_{missile})^2 V_2^2 =$$
$$= (M_{asteroid})^2 V_1^2 + (M_{missile})^2 V_{missile}^2. \quad (7.6)$$

and finally, solving for $V_2$,

$$\boxed{V_2 = \frac{\sqrt{(M_{asteroid})^2 V_1^2 + (M_{missile})^2 V_{missile}^2}}{M_{asteroid} + M_{missile}}}. \quad (7.7)$$

***This is the first, basic equation of the theory of orthogonal asteroid deflection: it yields the speed of the deflected asteroid (at the point of intercept) in terms of the incoming asteroid's mass and speed and of the deflecting missile's mass and speed.***

7) Having found $V_2$ from (7.7), it is easy to get the first of the three deflected asteroid's orbital elements, namely the semi-major axis $a_2$ of the deflected hyperbola. In fact, the energy integral of the deflected hyperbola at intercept reads

$$\frac{1}{2} V_2^2 - \frac{G M_{Earth}}{r_{intercept}} = \frac{G M_{Earth}}{2 a_2}. \quad (7.8)$$

Solving this for $a_2$ yields the ***deflected hyperbola's semi-major axis*** $a_2$:

$$\boxed{a_2 = \frac{G M_{Earth}}{V_2^2 - \frac{2 G M_{Earth}}{r_{intercept}}}}. \quad (7.9)$$

***This is the second, basic equation of the theory of orthogonal asteroid deflection: it yields the semi-major axis $a_2$ of the deflected asteroid's hyperbola in terms of the incoming asteroid's hyperbola elements and of the deflecting missile's mass and speed.***

8) The ***total energy of the deflected hyperbola***, $E_2$, follows at once from (7.9):

$$E_2 = \frac{G M_{Earth}}{2 a_2}. \quad (7.10)$$

9) Since the total energy of the deflected hyperbola is now known, one may write ***the total energy integral of the deflected hyperbola***:

$$\frac{1}{2} V_3^2 - \frac{G M_{Earth}}{r_3} = E_2. \quad (7.11)$$

This is an equation relating the asteroid's distance $r_3$ from the Earth to the corresponding asteroid's speed, $V_3$, at any time along the deflected hyperbola. It could thus more correctly written in the form

$$V_3(r) = \sqrt{2 E_2 + \frac{2 G M_{Earth}}{r}}. \quad (7.12)$$

Also, (7.11) may be the starting equation to compute a second deflection. And, in turn, this second deflection, if insufficient, could be followed by a third deflection. And so on and so forth we may shoot as many missiles as we need, and achieve what we call the ***"repeated deflection capability"***.

10) A glance to Figure 4 immediately yields the expression of the deflection angle, $\vartheta_{deflection}$, spanned by the deflected and undeflected asteroid's linear momentum vectors, $\vec{p}_2$ and $\vec{p}_1$, in the rectangular triangle :

$$\vartheta_{deflection} = \operatorname{atan}\left(\frac{p_{missile}}{p_1}\right). \quad (7.13)$$

11) From the same rectangular triangle, we are now going to derive the expression of the angle $\chi_1$ spanned by the incoming asteroid's linear momentum vector $\vec{p}_1$ and the radius vector to the asteroid at the interception point. We assume this angle to be positive when computed counterclockwise. In order to derive a formula for this angle $\chi_1$, we first notice that $\chi_1$ is the difference between two angles:

a) the angle ($\pi$ - phi_intercept) between the $x$ axis and the asteroid's radius vector at intercept, minus

b) the angle $\tau_1$ between the $x$ axis and the tangent vector to the incoming asteroid's path at intercept. To the express this angle $\tau_1$, we must firstly describe the tangent straight line to the incoming asteroid's hyperbola, that is the asteroid's hyperbola "just and instant before the missile hits the asteroid". Since the orientation of this tangent obviously changes according to where the tangency point is, in general its x and y components must be functions of the true anomaly $\phi$. Thus, upon differentiating the transformation equations

$$\begin{cases} x_{hyp}(\phi) = r_{hyp}(\phi)\cos(\phi) \\ y_{hyp}(\phi) = r_{hyp}(\phi)\sin(\phi) \end{cases} \quad (7.14)$$

with respect to the only independent variable $\phi$, one gets

$$\begin{cases} \dfrac{dx_{hyp}}{d\phi} = \dfrac{dr_{hyp}}{d\phi}\cos(\phi) - r_{hyp}\sin(\phi) \\ \dfrac{dy_{hyp}}{d\phi} = \dfrac{dr_{hyp}}{d\phi}\sin(\phi) + r_{hyp}\cos(\phi). \end{cases} \quad (7.15)$$

These are the components of the tangent vector to the incoming asteroid's hyperbola at any point along the hyperbola. The angle $\tau_1$ is clearly given by the arc tangent of the ratio of the two equations (7.15) computed at the point of intercept:

$$\tau_1 = \operatorname{atan}\left(\frac{\left.\dfrac{dy_{hyp}}{d\phi}\right|_{\phi=\pi}}{\left.\dfrac{dx_{hyp}}{d\phi}\right|_{\phi=\pi}}\right). \quad (7.16)$$

In conclusion, the angle $\chi_1$ is given by

$$\chi_1 = \left|(\pi - \text{phi\_intercept}) - \operatorname{atan}\left(\frac{\left.\dfrac{dy_{hyp}}{d\phi}\right|_{\phi=\pi}}{\left.\dfrac{dx_{hyp}}{d\phi}\right|_{\phi=\pi}}\right)\right|. \quad (7.17)$$

In the last formula we chose to introduce the absolute value of the whole right-hand side had to remind that the conventions about the signs of all angles have to be respected carefully while writing a computer code. Only in this case can numerical values to be avoided.

12) Just as for the angle $\chi_1$, so the angle $\chi_2$ can be computed in a very similar fashion. From Figure 4, it appears that the angle $\chi_2$ is, by definition, spanned by the outgoing asteroid's linear momentum vector $\vec{p}_2$ and the radius vector to the asteroid at the interception point. We assume this angle to be positive when computed counterclockwise. In order to derive a formula for this angle $\chi_2$, we first notice that $\chi_2$ is the difference between two angles:

c) the angle ($\pi$ - phi_intercept) between the $x$ axis and the asteroid's radius vector at intercept, minus

d) the angle $\tau_2$ between the $x$ axis and the tangent vector to the outgoing asteroid's path at intercept. To the express this angle $\tau_2$, we must firstly describe the tangent straight line to the outgoing asteroid's hyperbola, that is the asteroid's hyperbola "just an instant after the missile hits the asteroid". Since the orientation of this tangent obviously changes according to where the tangency point is, in general its x and y components must be functions of the true anomaly $\phi$. Thus, upon differentiating the transformation equations (where the subscript "hyp2" refers to the outgoing asteroid's hyperbola, that is the deflected asteroid's hyperbola)

$$\begin{cases} x_{hyp2}(\phi) = r_{hyp2}(\phi)\cos(\phi) \\ y_{hyp2}(\phi) = r_{hyp2}(\phi)\sin(\phi) \end{cases} \quad (7.18)$$

with respect to the only independent variable $\phi$, one gets

$$\begin{cases} \dfrac{dx_{hyp2}}{d\phi} = \dfrac{dr_{hyp2}}{d\phi}\cos(\phi) - r_{hyp2}\sin(\phi) \\ \dfrac{dy_{hyp2}}{d\phi} = \dfrac{dr_{hyp2}}{d\phi}\sin(\phi) + r_{hyp2}\cos(\phi). \end{cases} \quad (7.19)$$

These are the components of the tangent vector to the outgoing asteroid's hyperbola at any point along the hyperbola. The angle $\tau_2$ is clearly given by the arc tangent of the ratio of the two equations (7.19) computed at the point of intercept:

$$\tau_2 = \operatorname{atan}\left(\frac{\left.\frac{dy_{hyp2}}{d\phi}\right|_{\phi=\pi}}{\left.\frac{dx_{hyp2}}{d\phi}\right|_{\phi=\pi}}\right). \qquad (7.20)$$

In conclusion, the angle $\chi_2$ is given by

$$\chi_2 = \left|(\pi - \text{phi\_intercept}) - \operatorname{atan}\left(\frac{\left.\frac{dy_{hyp2}}{d\phi}\right|_{\phi=\pi}}{\left.\frac{dx_{hyp}2}{d\phi}\right|_{\phi=\pi}}\right)\right|. \qquad (7.21)$$

***This is the third, basic equation of the theory of orthogonal asteroid deflection: it yields the angle between the deflected asteroid's speed vector and the asteroid's radius vector at the interception point. It will play a key role in the subsequent calculation of the deflected asteroid's constant angular momentum and eccentricity.*** In the last formula we chose to introduce the absolute value of the whole right-hand side had to remind that the conventions about the signs of all angles have to be respected carefully while writing a computer code. Only in this case can numerical errors be avoided.

13) Having found the two angles $\chi_1$ and $\chi_2$, the way is now paved to find the two (different) angular momentum constants, denoted by $h_1$ and $h_2$, respectively, of each of the two hyperbolae under consideration. By definition, these two constants are defined by

$$h_1 = r_1 V_1 \sin \chi_1 \qquad (7.22)$$

and

$$\boxed{h_2 = r_1 V_2 \sin \chi_2}. \qquad (7.23)$$

***This is the fourth, basic equation of the theory of orthogonal asteroid deflection: it yields the deflected asteroid's constant angular momentum.*** Notice that, in (7.23), $r_1$ replaces the unnecessary $r_2$ since they are the same thing, i.e. the asteroid's radius vector at the interception point.

14) The next step is finding the eccentricity $e_2$ of the asteroid's deflected hyperbola by deriving it from the relevant angular momentum constant, $h_2$ (obviously, the corresponding eccentricity $e_1$ of the asteroid's incoming hyperbola is well known already!). As every textbook about the two body problem shows, the relevant equation reads

$$\boxed{e_2 = \sqrt{1 + \frac{h_2^2}{G M_{Earth} a_2}}}. \qquad (7.24)$$

***This is the fifth, basic equation of the theory of orthogonal asteroid deflection: it yields the eccentricity of the deflected asteroid's hyperbola (obviously in terms of the incoming asteroid's hyperbola and of the deflecting missile's mass and speed).***

15) Finally, the deflected hyperbola's argument of the perigee, $\omega_2$, remains to be found.

To determine $\omega_2$, consider the deflected hyperbola's polar equation

$$r(\phi) = \frac{a_2 \left(e_2^2 - 1\right)}{1 + e_2 \cos(\phi - \omega_2)}. \qquad (7.25)$$

In this equation, $a_2$ and $e_2$ are known already, while $\phi$ is the independent variable and $r$ is its function. If we want to change (7.25) into a single equation in the only unknown argument of the perigee, $\omega_2$, we must clearly assign precise values to both $\phi$ and $r$. These precise values must be the polar coordinates of the point of intercept, the only point shared by both the undeflected and the deflected asteroid's hyperbolae. By replacing these values into (7.25), one gets

$$r_1 = \frac{a_2 \left(e_2^2 - 1\right)}{1 + e_2 \cos(\phi_{intercept} - \omega_2)}. \qquad (7.26)$$

When one tries to solve (7.26) with respect to the only unknown $\omega_2$, one gets an expression with the arccos(…) function that must be handled carefully in order not to loose any root. Thus, by resorting to the same procedure already used for the arccos(…) function encountered in eqs. (4.1), (4.2) and (4.3), we firstly define a new constant $K$ given by

$$K = \frac{1}{e_2}\left[\frac{a_2 \left(e_2^2 - 1\right)}{r_1} - 1\right]. \qquad (7.27)$$

Then the argument of the perigee $\omega_2$, found by solving (7.26) has the two different values $\omega_2^{(1)}$ and $\omega_2^{(2)}$ given by writing down the expressions of the two resulting solutions of cos(…) = … with period $2\pi$:

$$\omega_2^{(1)} = 2\operatorname{atan}\left(\frac{\sqrt{1-K^2}+\sin(\phi_{intercept})}{K+\cos(\phi_{intercept})}\right)+m\cdot 2\pi \quad (7.28)$$

and

$$\boxed{\omega_2^{(2)} = -2\operatorname{atan}\left(\frac{\sqrt{1-K^2}-\sin(\phi_{intercept})}{K+\cos(\phi_{intercept})}\right)+n\cdot 2\pi} \quad (7.29)$$

It is not surprising that (7.28) and (7.29) (where $m, n = 1, 2, 3...$) yield two different arguments of the perigee, denoted $\omega_2^{(1)}$ and $\omega_2^{(2)}$, for the deflected asteroid's hyperbola. In fact, any hyperbola with its focus centered at the origin (the Earth) may only intersect in *two* different ways an assigned, single point: by either one or the other one of its two legs departing from the perigee in the two opposite directions. And the root to use in this case is given by (7.29) because this is the only angle larger than 90 degrees and smaller than 180 degrees, as expected. Thus, (7.29) *is the sixth, basic equation of the theory of orthogonal asteroid deflection: it yields the argument of the perigee of the deflected asteroid's hyperbola (obviously in terms of the incoming asteroid's hyperbola and of the deflecting missile's mass and speed).*

16) In conclusion, *the polar equation of the deflected asteroid's hyperbola is given by*

$$\boxed{r(\phi)=\frac{a_2\left(e_2^2-1\right)}{1+e_2\cos\left(\phi-\omega_2^{(2)}\right)}} \quad (7.30)$$

*The problem of determining the deflected asteroid's hyperbola is thus completely solved.*

## 8. ASTEROID'S ORTHOGONAL DEFLECTION: ITS CORRECT GRAPHICAL REPRESENTATION

The numerical example of asteroid deflection shown in Figure 4 (and computed by virtue of the code written by this author) was ***grossly exaggerated*** to let the reader fully appreciate the rectangular triangle of the deflection and the deflected hyperbola stemming out of it.

The correct graphical representation of both the undeflected and deflected hyperbolae, however, reveals quite a different picture. This correct (i.e. to scale) graphical representation is shown in Figure 5 for the particular numerical case characterized by (no argument now about how "reasonable" the assumed numerical assumptions might be, please!):

1) An assumed incoming asteroid's hyperbola with a semi-major axis of 50000 km;
2) An assumed incoming asteroid's hyperbola with an eccentricity of 2;
3) An assumed incoming asteroid's hyperbola with an argument of the perigee of $\pi/4$;
4) An assumed incoming asteroid's hyperbola with a perigee distance of 50000 km (this is the minimal distance from the Earth center that the asteroid would reach were there *no* deflection; also, this incoming asteroid's hyperbola is characterized by an "impact parameter" ( = orthogonal distance from the Earth at infinity) of 86602 km);
5) An assumed asteroid diameter of 1 km;
6) An assumed asteroid's average density of 4.135 x $10^3$ kg/m$^3$ (this is the estimated average density of asteroid 4 Vesta), yielding a total asteroid mass of 216 x $10^{10}$ kg;
7) An assumed incoming asteroid speed of 3.14 km/sec along the incoming hyperbola at the point of intercept by the missile shot from L3;
8) An assumed missile mass of 211 tonnes (i.e. a missile of the last "SS 18" ex-Soviet ICBM class);
9) An assumed missile speed of 1 km/sec at the point of intercept along the missile's ellipse shot from L3;
10) An assumed missile extra-boost of 0.5 km/sec to enhance the deflection by firing the missile engines "at the very last moment" before the collision;

Under these ten assumptions, the results of our simulation code yield the actual deflection shown in Figure 5. One immediately notices that:

a) The deflected asteroid's hyperbola (solid black line in Figure 5) can hardly be distinguished from the incoming asteroid's hyperbola (dashed line in the same Figure 5) in the vicinity of the point of intercept. Only at about 100000km from the Earth do the two hyperbolae become distinguishable and depart then more and more from one another;
b) The key goal of the whole Planetary Defense by Asteroid Deflection is, of course, increasing the asteroid's perigee distance. Our simulation yields the shown increase of the perigee from 50000 km to 62470 km. Not much. But here the true goal of all our calculations becomes apparent, that is…
c) *Find out the best trade-offs between asteroid's mass and speed, and deflecting missile's mass and speed to achieve the Optimal Confocal Planetary Defense.*

*This we leave to one more, forthcoming paper of ours: this paper is just a beginning !*

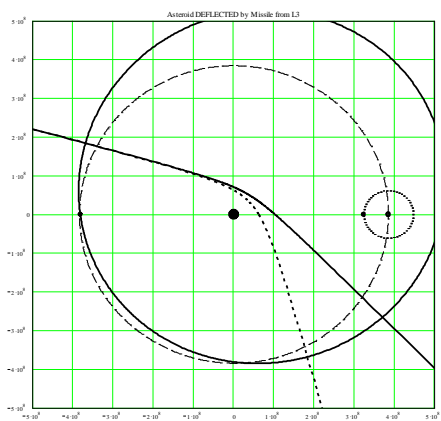

**Figure 5. Correct Graphical Representation of the grossly exaggerated Deflection shown in Figure 4. The two hyperbolae are hardly distinguishable until about 100000 km from the Earth. Then they depart from one another, and, in conclusion, the perigee is raised from 50000 km to 62470 km.**

## CONCLUSION

**Missiles to deflect asteroids are better shot from the Lagrangian Points L1 and L3 of the Earth-Moon system than from the Earth. They can then deflect asteroids at best by hitting them at an angle of just 90°. We have cast all this into the simple language of a Keplerian mathematical theory. A patented code along these lines, dubbed AsterOFF, was written (in a MathCad environment) to let many simulations be made.**

## REFERENCES

[1] C. Maccone, *"Planetary Defense from the Nearest 4 Lagrangian Points Plus RFI-Free Radioastronomy from the Farside of the Moon: A Unified Vision"*, Academy Transactions Note, Acta Astronautica, Vol. 50, (2002), pp. 185-199.
[2] C. Maccone, *"Deflecting Asteroids at 90°"*, a poster paper presented at the "NASA Workshop on Scientific Requirements for Mitigation of Hazardous Comets and Asteroids" held in Arlington, VA, September 3-6, 2002. This paper is as an Extended Abstract at page 70 of the long (8.7 MB) pdf file that may be downloaded from the web site http://www.noao.edu/meetings/mitigation/eav.html
[3] C. Maccone, *"Optimal Trajectories from the Earth-Moon L1 and L3 Points to Deflect Hazardous Asteroids and Comets"*, a paper presented at the "New Trends in Astrodynamics Conference", run by Ed Belbruno of Princeton University and held at the University of Maryland at College Park, January 20-22, 2003. The relevant Proceedings are under publication by the New York Academy of Sciences.
[4] D. W. Dunham, *"Utilization of libration-point orbits"*, in the book titled *"Problems of Astronautics and Celestial Mechanics"*, Proceedings of an International Symposium held at Politecnico di Torino, 10-12 June 1987, in honor of the late Prof. Giuseppe Colombo. Published as Supplement to Vol. 122 of the "Atti dell'Accademia delle Scienze di Torino" by the Academy of Sciences of Turin, 1990, p. 239-274.
[5] J. L. Lagrange, *Oeuvres* (M. A. J. Serret, editor), Vol. 6, Gauthier-Villars, Paris, 1873. Lagrange's main article on the three-body problem was dated 1772, but was not originally published until 1777 by Panekoucke, Paris; see page 304 of ref. [4].